\documentclass[journal]{IEEEtran}
\IEEEoverridecommandlockouts
\usepackage{amsmath,amssymb,amsfonts}
\usepackage{graphicx}
\usepackage{textcomp}
\usepackage{xcolor}
\usepackage{algorithm}
\usepackage{algpseudocode}

\usepackage[
backend=biber,
style=ieee,
sorting=none
]{biblatex}
\begin{document}

\title{Area Efficient Modular Reduction in Hardware for Arbitrary Static Moduli}

\author{\IEEEauthorblockN{Robin Müller, Willi Meier, Christoph F. Wildfeuer}\\
\IEEEauthorblockA{\textit{FHNW University of Applied Sciences and Arts Northwestern Switzerland} \\
\textit{School of Engineering}\\
5210 Windisch, Switzerland \\
robin.mueller@fhnw.ch}

}

\maketitle

\begin{abstract}
Modular reduction is a crucial operation in many post-quantum cryptographic schemes \cite{pqc_overview, pqc_challenges}, including the Kyber key exchange method \cite{original_kyber_paper} or Dilithium signature scheme \cite{dilithium}. However, it can be computationally expensive and pose a performance bottleneck in hardware implementations.
To address this issue, we propose a novel approach for computing modular reduction efficiently in hardware for arbitrary static moduli. Unlike other commonly used methods such as Barrett or Montgomery reduction, the method does not require any multiplications. It is not dependent on properties of any particular choice of modulus for good performance and low area consumption. 
Its major strength lies in its low area consumption, which was reduced by 60\% for optimized and up to 90\% for generic Barrett implementations for Kyber and Dilithium. Additionally, it is well suited for parallelization and pipelining and scales linearly in hardware resource consumption with increasing operation width. All operations can be performed in the bit-width of the modulus, rather than the size of the number being reduced. This shortens carry chains and allows for faster clocking. Moreover, our method can be executed in constant time, which is essential for cryptography applications where timing attacks can be used to obtain information about the secret key.
\end{abstract}

\begin{IEEEkeywords}
Modular Reduction, FPGA, Digital Hardware, Cryptography
\end{IEEEkeywords}

\section{Introduction}
Computation of the modulo operation is an important part of many cryptographic schemes, where operations are performed in a finite field. Among these are multiple candidates for post-quantum cryptographic schemes \cite{pqc_overview}, such as Kyber (key encapsulation) \cite{original_kyber_paper} or Dilithium (signature) \cite{dilithium}. Kyber is an IND-CCA2-secure key encapsulation mechanism, whose security is based on the hardness of solving the learning-with-errors problem over module lattices. The computation of the mod operation is among the most often performed operations in these schemes. It has to be performed many times after other arithmetic operations, such as additions or multiplications. Computing modular reductions efficiently can therefore help improve the overall performance and efficiency of these cryptographic schemes. There are multiple common techniques for computing the reduction in hardware, such as Barrett reduction \cite{barrett_implementing_1987} or Montgomery reduction \cite{montgomery_modular_1985}. A major advantage of these schemes, for software implementations, is that they require only few steps in sequential execution. In the case of hardware implementations, these schemes are still relatively expensive, since they rely heavily on multiplications. These consume significantly more chip area compared to other operations, such as additions. A commonly used technique to make these methods more hardware friendly is to have multiplication constants that contain relatively few ones in binary representation, making multiplication with them feasible by a few shifts and add operations. For this technique to be feasible, one needs to rely on the underlying scheme to use moduli for which Barrett constants can be constructed that contain only few binary ones. This makes the method susceptible to different choices of the modulus. In this paper, we show a method of performing modular reduction for a general modulus without using multiplication, using only a mixture of additions, look-up-tables and multiplexers. The implementation shows similar or smaller hardware use and performance on FPGAs as a number optimized Barrett approach, but without relying on any specific choice of the modulus.

The underlying idea of using various forms of precomputed values for reduction has been considered before \cite{cryptoeprint:2014/040, lim_fast_nodate, cryptoeprint:2014/755}. Most often, these attempts focus on efficient software implementation, which can differ significantly from efficient hardware implementations. For hardware implementations, a similar approach to ours has been shown in \cite{r_omondi_modular_2020}, where the underlying idea was used to form an iterative scheme. In this article, we present an optimized parallelized approach to modular reduction using precomputed values. It reduces area consumption compared to the Barrett method and increases data throughput compared to the related iterative approach \cite{r_omondi_modular_2020}.

In Section \ref{sec:barrett} we first introduce the commonly used Barrett reduction which we use as a reference to compare our scheme. We then introduce our method in \ref{sec:Algorithm} and discuss the underlying idea as well as the intended hardware implementation architecture. Finally, we show the implementation results on an FPGA platform and perform ASIC synthesis. We compare results for both Barrett and our approach in \ref{sec:impl_results}.

\section{Barrett Reduction}
\label{sec:barrett}
In this article, we compare our scheme to the commonly used Barrett reduction \cite{barrett_implementing_1987} method. We therefore briefly introduce the Barrett reduction as well as a commonly made optimization to it for hardware implementations. The main scheme discussed in this article (stating from \ref{sec:Algorithm}), does not rely on the Barrett reduction.
\subsection{General Barrett Reduction}
\label{sec:barret_general}
In this description of the Barrett reduction, we let $\Tilde{c}$ be a number that we want to reduce $\mod q$ to obtain the reduced number $c$, where $c \gets \Tilde{c} \mod q$. We do this by subtracting the correct multiple of the modulus $q$ from $\Tilde{c}$ to finally receive our result $c$ where $c<q$. To find the correct multiple of $q$, we divide by a power of $2$, which can simply be performed in hardware using shift operations. We store a constant $m$ which approximates the ratio between the performed power of two division and the modulus $q$. By multiplying the division result by $m$, we obtain the (almost) correct multiple of $q$ that needs to be subtracted. In Alg.~\ref{alg:barrett_general} we show Barrett reduction in the general case, which was taken from \cite{banerjee2019sapphire}.
\begin{algorithm}
\caption{General Barrett Reduction mod $q$ \cite{banerjee2019sapphire}}
\label{alg:barrett_general}
\begin{algorithmic}[1]
    \Require $\Tilde{c}\in [0, q^2)$, $m$ and $k$ such that $m=\lfloor2^k/q\rfloor$
    \Ensure $c = \Tilde{c} \mod q$
    \State $t \gets (\Tilde{c} \cdot m) \gg k$
    \State $c \gets \Tilde{c} - (t \cdot q)$
    \If{$c \geq q$}
        \State $c \gets c-q$
    \EndIf
    \State \Return $c$
\end{algorithmic}
\end{algorithm}

\subsection{Modulus Optimized Barrett Reduction}
\label{sec:barrett_optimized}
To optimize the Barrett reduction for better hardware resource efficiency, we mainly need to look at the multiplications in the scheme. Let $a, b$ be two factors that we want to multiply. We let $N$ be the index (starting from $0$) of the most significant bit in the larger of the two factors ($a$ or $b$) as defined in \eqref{eq:bit_nb_generic_mult}. We can compute the multiplication of $a$ times $b$ in binary representation, as shown in \eqref{eq:binary_mult}. For this, we shift $a$ by all integer values starting from $0$ to $N$ and sum those terms, where $b$ contains a binary one at the index by which we shifted $a$ (binary multiplication by either $1$ or $0$ in \eqref{eq:binary_mult}):
\begin{equation}
    N +1  = \lceil\log_2{\max(\{a, b\})}\rceil
    \label{eq:bit_nb_generic_mult}
\end{equation}
\begin{equation}
    a\cdot b = \sum_{i=0}^{N}(b_i \cdot (a_i \ll i))
    \label{eq:binary_mult}
\end{equation}
If in \eqref{eq:binary_mult} the binary representation of $b$ only contains a few ones, the multiplication with $b$ becomes much simpler, since only a few shifted terms of $a$ need to be considered in the sum. All terms where $b_i$ is zero can be omitted.  We can utilize this property to build efficient multipliers in cases where $a$ is a variable term but $b$ is constant. We now want to choose $b$ such that it contains as few binary ones as possible. In the case of Barrett reduction as introduced in Alg.~\ref{alg:barrett_general}, those constant terms by which we need to multiply, would be $q$ as well as $m$. Alg.~\ref{alg:optimKyber} shows how this scheme would be applied to Barrett reduction in the case of the Kyber Post Quantum Cryptography algorithm \cite{computer_security_division_round_2017}, where $q=3329$. In this case, the multiplication by $q$ benefits heavily from this optimization, since 3329 (binary: $110100000001$) only contains 4 binary ones. Multiplication by $m$ only benefits from it to a limited extent, since it contains 9 binary ones. While this form of optimization can lead to highly efficient modular reduction in some cases of $q$ and $m$, the gained resource efficiency through optimization, can vary significantly between different choices of $q$ and $m$. Furthermore, the value of $q$ (and $m$ indirectly since it depends on $q$) is most often given by the cryptographic scheme and can often not be chosen solely on the criteria of efficient Barrett reduction.

\begin{algorithm}
\caption{Barrett mod 3329}\label{alg:optimKyber}
\begin{algorithmic}[1]
\Require $q = 2^{11} + 2^{10} + 2^8 + 1, m=5039, k=24, \Tilde{c}\in [0, q^2) $
\Ensure $c = \Tilde{c} \mod q$
\State $t \gets (\Tilde{c}\ll 12)+(\Tilde{c}\ll 9)+(\Tilde{c}\ll 8)+(\Tilde{c}\ll7)+ (\Tilde{c}\ll5)+$
    \State $(\Tilde{c}\ll3)+(\Tilde{c}\ll 2)+(\Tilde{c}\ll 1)+\Tilde{c}$ \Comment{Equiv. $m\cdot \Tilde{c}$}
\State $t \gets t \gg k$
\State $t \gets (t\ll 11)+(t \ll 10)+(t \ll 8)+t$ \Comment{Equiv. $q \cdot t$}
\State $c \gets \Tilde{c} - t$
\If{$c \geq q$}
    \State $c \gets c-q$
\EndIf
\State \Return $c$
\end{algorithmic}
\end{algorithm}

\section{Algorithm}
\label{sec:Algorithm}
In this section, we show the proposed method for computing modular reduction and elaborate the hardware implementation of the scheme.

\subsection{Foundation}
We will consider the case that we want to perform operations (addition or multiplication) within a ring of positive integers modulo $q$, which from here on we denote as $\mathbb{Z}_q$. Let $a, b \in \mathbb{Z}_q$. We define $c \in \mathbb{Z}_q$ as the result of ($a+b$) or ($a\cdot b$) after reduction (mod $q$) and $\Tilde{c}$ as the operation result before reduction. For the case of addition, we can see in \eqref{eq:add_case}, that the maximum number we will obtain from adding $a$ and $b$ is smaller than $2q$. This can be seen by adding the two largest possible elements in $\mathbb{Z}_q$ being $q-1$. We can therefore simply compute $(c \gets \Tilde{c} \mod q)$ from $\Tilde{c}$ by a conditional subtraction of $q$, as in \eqref{eq:cond_subtract}.
\begin{equation}
    \forall a,b\in\mathbb{Z}_q: \max{\Tilde{c}}=\max(a+b)=2(q - 1) \label{eq:add_case}
\end{equation}
\begin{equation}
    c = (\Tilde{c} \mod{q}) = \begin{cases}
         \Tilde{c} & \Tilde{c} < q  \\
         \Tilde{c}-q &  \Tilde{c} \geq q
         \label{eq:cond_subtract}
    \end{cases}
\end{equation}
For multiplication \eqref{eq:mult_case}, the reduction becomes significantly more complex to compute as our range of results becomes significantly larger. If we wanted to reduce $\Tilde{c}$ to $c$ we would have $q-1$ cases to consider, rendering a solution like in \eqref{eq:cond_subtract} impractical. 
\begin{equation}
    \forall a,b\in\mathbb{Z}_q: \max{\Tilde{c}}=\max{(a\cdot b)}=(q-1)^2
    \label{eq:mult_case}
\end{equation}
To solve the issue of large intermediate results, we want to consider the following property:
Let $x,y \in \mathbb{Z}$. We can see from \eqref{eq:equivalence_mod}, that when computing a sum (mod $q$), we can also first compute the reduction of the summands individually, before adding the individual elements together and reducing the resulting sum again.
\begin{equation}
    x+y \equiv (x \mod{q}) + (y\mod{q})  \pmod{q}
    \label{eq:equivalence_mod}
\end{equation}

This property can be useful for bringing the sum closer to elements in $\mathbb{Z}_q$, such that the final reduction can be performed as in \eqref{eq:cond_subtract}. In such a case, as the final reduction of the sum, we only need to conditionally subtract $q$ if the sum is $\geq q$, since the sum of two elements in $\mathbb{Z}_q$ will be at most $2(q - 1)$. For this property to be useful, we additionally need the steps $(x \mod{q})$ and $(y\mod{q})$ to be easy to perform. We will see in the following section how we can express a multiplication result as in \eqref{eq:mult_case} in such a form.

\subsection{Reduction Scheme}
\label{sec:algorithm_basic}
We define $\Tilde{c}$ as the intermediate result, before reduction $(c = \Tilde{c} \mod q)$, of multiplying $(a\cdot b)$ with $ a,b \in \mathbb{Z}_q$. We can interpret $\Tilde{c}$ as a $2n\text{-bit}$ vector, where $n= \lceil\log_2{q}\rceil$ is the number of bits needed to store an element in $\mathbb{Z}_q$. We make the difference between interpreting $\Tilde{c}$ as an integer or bit vector explicit by using bold $\mathbf{\Tilde{c}}$ for the bit vector interpretation and normal font $\Tilde{c}$, as the integer interpretation. The relationship between $\Tilde{c}$ and $\mathbf{\Tilde{c}}$ is defined as follows:

\begin{equation}
    \mathbf{\Tilde{c}} = [\Tilde{c}_0, \Tilde{c}_1, ..., \Tilde{c}_{2n - 1}],\,  \Tilde{c}_i \in \{0,1\}
    \label{eq:bit_vec}
\end{equation}
\begin{equation*}
    \Tilde{c} = \sum_{i=0}^{2n-1} (\Tilde{c}_i \cdot 2^i)
\end{equation*}
As introduced in \eqref{eq:equivalence_mod}, in a sum, we can reduce summands individually, before reducing the resulting sum. By writing $\Tilde{c}$ as a bit vector $\mathbf{\Tilde{c}}$, we can express the potentially large (relative to $q$) number $\Tilde{c}$, as a sum of powers of two. When applying the idea of reducing summands individually from \eqref{eq:equivalence_mod}, to $\mathbf{\Tilde{c}}$, we end up with \eqref{eq:equiv_sum}:
\begin{equation}
    c = \sum_{i=0}^{2n-1} (\Tilde{c}_i \cdot 2^i \mod{q}) \pmod{q} \label{eq:equiv_sum}
\end{equation}

As previously stated, for the summation property \eqref{eq:equivalence_mod} to be useful, we need to be able to easily compute $(x \mod q)$ for a summand $x$. We can observe that for any given bit index $i$, the value of $(2^i \mod q)$ is constant, since both $q$ and $2^i$ are constant. These values are independent of the bit-values in $\mathbf{\Tilde{c}}$, we only changed the value of how we interpret a bit $\Tilde{c}_i$ in $\mathbf{\Tilde{c}}$ being one from $2^i$ to  $(2^i \mod q)$ when forming the integer representation from the bit vector representation. Summing the new bit values, results in the sum of them being congruent to the original value of $\Tilde{c}$. The difference is, that the maximum value is now much closer to $q$, since we are summing bits whose values are elements in $\mathbb{Z}_q$. We can easily precompute and store the values of  $(2^i \mod q)$, making the condition we've set, that it needs to be easy to obtain $(x \mod q)$ in \eqref{eq:equivalence_mod} true. We can rewrite an individual summand in \eqref{eq:equiv_sum} to more clearly separate the variable bit-value from the constant part of the expression: 
\begin{equation}
(\Tilde{c}_i \cdot 2^i \mod{q}) = \begin{cases}
        (2^i \mod{q}) & \Tilde{c}_i = 1  \\
         0 &  \Tilde{c}_i = 0
         \end{cases}
\end{equation}
Storing the values of $(2^i \mod{q})$ for all $i$ in a pre-computed table would require us to store a total $2n$ $n\text{-bit}$ values. We can further simplify this to just storing $n+1$ values, since the $n$ least significant bits are already elements in $\mathbb{Z}_q$:
\begin{equation}
\forall i < n : (2^i \mod{q}) = 2^i
\end{equation}
We can further group the $n-1$ least significant bits to form a single summand since their sum will always be an element in $\mathbb{Z}_q$:
\begin{equation}
q < \sum_{i=0}^{n-2} 2^i
\label{eq:third_property}
\end{equation}
Combining the things discussed so far in this section, we want to compute $c$ from the bit vector $\mathbf{\Tilde{c}}$ as follows in \eqref{eq:combined_calc}. The left sum is marked gray to point out the difference, that the left sum just requires interpreting a bit-vector in the classical way as an integer, without needing any actual computation. On the contrary, the right sum actually involves summing terms of $n\text{-bit}$ numbers. Note that we deliberately include also the $(n-1)\text{th}$ bit in the left sum, allowing it to get slightly larger than $q$ (but smaller than $2q$) which we need to (and easily can) account for in the final (mod $q$) step.
\begin{equation}
    c = \left (\textcolor{gray}{ \sum_{i=0}^{n-1} (\Tilde{c}_i \cdot 2^i)} + \sum_{i=n}^{2n-1} (\Tilde{c}_i \cdot (2^i \mod{q})) \right ) \mod q
    \label{eq:combined_calc}
\end{equation}
The final thing left up for discussion is how to compute the final reduction (mod $q$). Thanks to the term before the final reduction just being a sum of elements in $\mathbb{Z}_q$, we can again look at the largest number we obtain from summing two elements in $\mathbb{Z}_q$ \eqref{eq:add_case}. Since the largest value that an individual summand can have is $(q-1)$ and we are summing $n$ such terms, the maximum number that can occur will therefore be $n(q-1)$. Since we previously allowed the term of least significant bits to become slightly larger than, $q$ we need to account for this here by letting the maximum number we want to reduce be $(n+1) \cdot (q-1)$. Since there are only $n + 1$ cases to consider, we can relatively cheaply build a compare logic to subtract the correct multiple of $q$ from the intermediate result to compute the final reduction step.
\begin{algorithm}[H]
    \caption{LUT based reduction scheme}\label{alg:lut_scheme}
    \begin{algorithmic}[1]
    \Require $f[i] = 2^i \mod q$, $n = \lceil\log_2{q}\rceil$, $N = 2n$, $\Tilde{c}\in [0, 2^N)$
    \Ensure $c = \Tilde{c} \mod q$
    \State $c \gets\Tilde{c} \mod 2^n$ \Comment{Take lower $n$ bits}
    \State $c \gets c + \sum_{i=n}^{N-1}(\Tilde{c_i} \cdot f[i])$
    \While{$c \geq q$} \Comment{Maximum of $n+1$ iterations}
        \State $c \gets c - q$
    \EndWhile
    \State \Return c
    \end{algorithmic}
\end{algorithm}
Alg.~\ref{alg:lut_scheme} combines all the steps described in this section. It should be pointed out that the subtraction of $q$ in a while loop is not intended to be implemented this way, since it would neither be very efficient nor possess a constant execution time. In the next section, we are going to discuss how we can put this scheme efficiently into hardware and further optimize the summation step and the final subtraction of multiples of $q$.

\subsection{Optimization for Hardware}
\label{sec:Algo_Optim_HW}
\begin{figure}[htbp]
\centerline{\includegraphics[width=\linewidth]{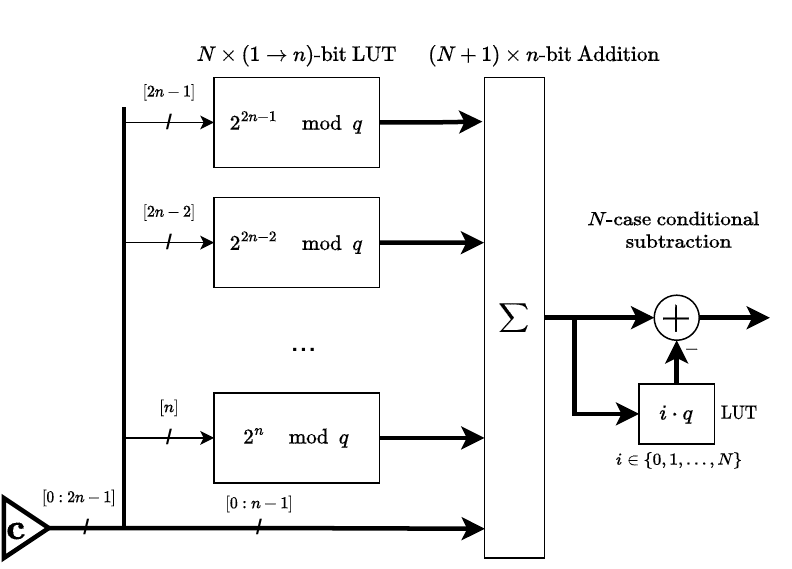}}
\caption{The basic implementation of the proposed method before hardware optimization}
\label{fig:basic_method}
\end{figure}

\begin{figure}[htbp]
\centerline{\includegraphics[width=\linewidth]{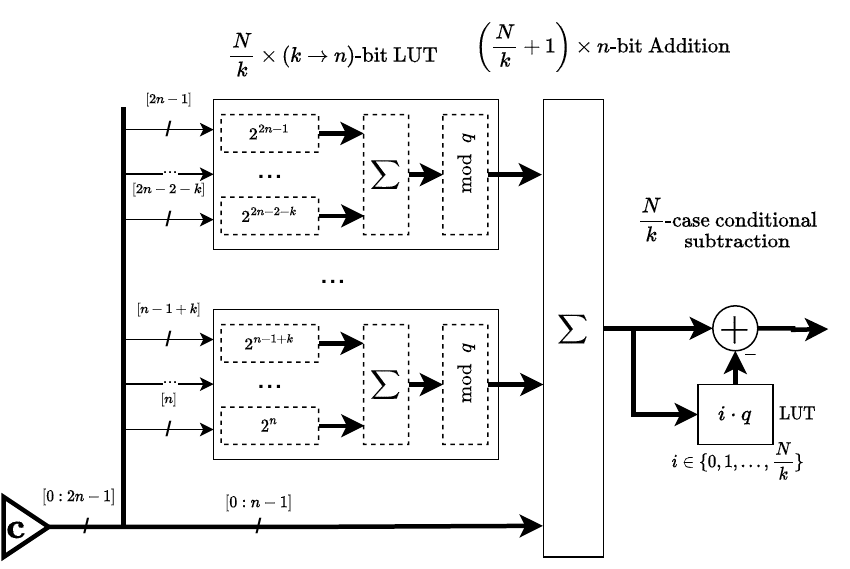}}
\caption{The optimized implementation, forming a more balanced tradeoff of storage and arithmetic resources. Dashed lines indicate the logic encoded inside the LUT. The values are pre-computed and stored in the LUT for all possible $2^k$ input states of the table.}
\label{fig:optimized_method}
\end{figure}

From the method discussed in section \ref{sec:algorithm_basic} we can build a hardware structure as shown in Fig.~\ref{fig:basic_method} that resembles this functionality. We introduce $N$ as the number of bits that need to be fed through a look-up table, which is every bit above the $n\text{th}$ bit in the input bit-vector. In the case of reduction after the multiplication of two elements in $\mathbb{Z}_q$, the resulting bit length will be $2n$, $N=2n-n=n$. The circuit in Fig.~\ref{fig:basic_method} will require only a small number of pre-computed values for the first Look-up-Table (LUT) stage ($N$ $n\text{-bit}$ values). However, the implementation will require $N+1$ $n\text{-bit}$ additions to sum up the table outputs. Additionally, in the final reduction step, which consists of subtracting the correct multiple of $q$ to fulfil $c<q$, all multiples of $q$ from $0\cdot q\text{ to }N\cdot q$ need to be considered.
For a more efficient hardware implementation, we have two main goals with the previously introduced basic method from Fig.~\ref{fig:basic_method}. The first one is to reduce the number of cases needed to be considered for conditional subtraction (final reduction step). The second one is to reduce the number of additions needed to sum the table outputs. What we therefore do is group our previous $1\text{-bit}$ input LUTs from Fig.~\ref{fig:basic_method} into larger LUTs of $k\text{-bit inputs}$ as illustrated in Fig.~\ref{fig:optimized_method}. $k$ is a freely choosable configuration parameter, that can be adapted to best fit a target device's architecture (e.g. available LUT size of an FPGA). Note that table sizes don't need to be symmetric, 
 however for better demonstration of the effect of grouping tables together, we here chose them to be all $k\text{-bit}$ input tables. By grouping single input bit tables into larger $k\text{-bit input}$ tables, the number of precomputed values we need to store increases by a factor $2^{k-1}$. The advantage on the other side is that the number of additions we needed in the second stage decreases by (almost) a factor of $k$, since fewer total LUT outputs need to be summed. This in turn reduces the number of cases (different multiples of $q$) that need to be considered for the final reduction step by (almost) a factor of $k$. The trick to reducing the number of possible subtraction cases in the final step is, that we do not simply group and sum all table configurations. We instead also compute a modular reduction on all table entries, turning sums of elements in $\mathbb{Z}_q$, which could normally get larger than $q$, back into elements in $\mathbb{Z}_q$. Bringing the new maximum of the intermediate sum closer to $q$ (from $(N+1)\cdot(q-1)$ to $(\frac{N}{k}+1)\cdot(q-1)$).

The following equation adapts the original equation describing the basic variant of the scheme \eqref{eq:combined_calc} to the hardware optimized variant shown as a circuit in Fig.~\ref{fig:optimized_method}:
\begin{equation}
    \hat{c}= \sum_{i=0}^{n-1} (\Tilde{c}_i \cdot 2^i) +\sum_{i=0}^{\frac{N}{k}-1}\left (\sum_{j=0}^{k-1}(\Tilde{c}_{(n+j+ki)} \cdot 2^{n+j+ki})\mod q\right)
\end{equation}
\begin{equation*}
    c = \hat{c} - i q  \, | \, i q \leq \hat{c} < (i +1)q \, \land \, i \in [0, N/k]
\end{equation*}

\subsection{Optimizing the Final Reduction Step}
To optimize the final reduction step, we need to minimize the maximum value of the intermediate sum $\hat{c}$. The maximum value is determined by the maximum value of the term not feed through the LUTs $(2^n - 1)$ (the value of all bits in this term being one) and the sum of the largest entries in the tables $f_i[x]$:
\begin{equation}
    \max(\hat{c}) = (2^n - 1) + \sum_{i = 0}^{\frac{N}{k}-1} \max(f_i[x]) \approx (2^n - 1) + q\frac{N}{k}
\end{equation}
In most cases, at least one combination of the input bit vector of a table leads to a table entry that is close to the value of $q$. We can thereby approximate the maximum of a table as $\max(f_i[x]) \approx q$. 

One thing we can do to reduce the maximum value of $\max(\hat{c})$ that is often worth doing, is including the $n$th bit in the tables. It could theoretically also be forwarded to $\hat{c}$ directly, as the mod operation does not change its value ($2^{n-1} = (2 ^{n-1} \mod q)$). It can however make sense to include this bit in one of the tables. Removing the most significant bit from the term not feed through the LUTs assures that its value remains much smaller than $q$. This often leads to having to consider one less case in the final reduction step.

Another optimization regards how we group the input bits into tables. The case where this is relevant is, when the maximum value of $\hat{c}$ is just above an integer threshold (e.g. $\max(\hat{c}) = 3.05$, anything $<3$ would allow us to only need to check for multiples of $q$ up to 2 instead of 3). In this case, we would like to lower $\max(\hat{c})$ slightly below the integer threshold to remove one case from the final reduction step. To achieve this, we can rearrange how we group the input bit vector $\mathbf{\Tilde{c}}$ into tables. We have so far considered the bits to be grouped in their natural order (e.g. grouping $[c_1,c_2, c_3,c_4]$ into 2 tables of two inputs each would be $[[c_1,c_2],[c_3, c_4]]$). We can often slightly affect the maximum value $\max(\hat{c})$ by  choosing a different configuration (e.g. $[[c_1,c_3],[c_2, c_4]]$). We utilize here that $\max(f_i[x])$ is only approximately equal to $q$. We can vary the exact value $\max(f_i[x])$ by varying which input bits we group together, such that the sum of $\max(f_i[x])$ becomes as small as possible.

\subsection{Numeric Example}
\label{sec:numeric_example}
We are now going to apply the reduction scheme introduced in \ref{sec:algorithm_basic} and its hardware optimized variant from \ref{sec:Algo_Optim_HW}, in a simple numeric example with only small bit vector sizes. We let the modulus $q=13$ and the corresponding number of bits for an element in $\mathbb{Z}_q$ $n=4$. We want to reduce an arbitrarily selected $2n = 8$ bit long number $\Tilde{c} = 210$ (binary: $11010010$) to $c = \Tilde{c} \mod q = 210 \mod 13 = 2$. Since our modulus $q$ requires four bits, we leave the four lower bits ($\Tilde{c}_0$ to $\Tilde{c}_3$) untouched and precompute four single input tables (table values $0$ and $(2^i \mod 13)$) for the four upper bits ($\Tilde{c}_4$ to $\Tilde{c}_7$):
\begin{table}[H]
    \centering
    \begin{tabular}{c|c}
        $i$ &  $2^i \mod 13$\\
        \hline
        $4$ & $3$ \\
        $5$ & $6$\\
        $6$ & $12$\\
        $7$ & $11$\\
    \end{tabular}
    \caption{Table for the basic reduction scheme with single input tables}
    \label{tab:single_input_table_example}
\end{table}
To design the final reduction step, we consider what the largest number $\hat{c}$ would be. This would, for the basic reduction scheme, be the case if all bits in $\mathbf{\Tilde{c}}$ are $1$ and their values would be summed up:
\begin{equation}
    \max{(\hat{c})} = 1 + 2 + 4+ 8+ 3+ 6+ 12+ 11 = 47
\end{equation}
\begin{equation}
    i_{\max} = \frac{\max{\hat{c}}}{q}=\frac{47}{13} \approx 3.6
    \label{eq:max_non_grouped_reduction}
\end{equation}
\begin{equation}
     c = \hat{c} - i q  \, | \, i \in [0, \lfloor 3.6 \rfloor]
\end{equation}
In this case, our final reduction step would need to consider all multiples of $q$ from $0$ to $3$. We now have determined all needed parameters for an instance of the reduction hardware and can start reducing numbers at runtime.
We now apply $\Tilde{c} = 210$ to the input of the circuit. The lower four bits will be forwarded into the intermediate sum unchanged, which will form the first summation term $0010 \rightarrow 2$. In the upper four bits, the tables will contribute a $0$ to the sum if the input was also $0$, or its precomputed value if the input was $1$. Since $\Tilde{c}_4$, $\Tilde{c}_6$ and $\Tilde{c}_7$ where $1$ in the upper four bits, we add their corresponding value from Table~\ref{tab:single_input_table_example}, which would be: $\Tilde{c}_4 \rightarrow 3$, $\Tilde{c}_6 \rightarrow 12$ and $\Tilde{c}_7 \rightarrow 11$. The rest of the summation terms will be zero. We therefore obtain the intermediate sum:
\begin{equation}
    \hat{c} = 2 + 3 +12+ 11 = 28
\end{equation}
We now check which multiple of $q$ we need to subtract in the final step. We can then perform the final subtraction and obtain the result:
\begin{equation}
     2q \leq \hat{c} < 3q \rightarrow i = 2
\end{equation}
\begin{equation}
    c = \hat{c} - (i\ \cdot q) = 28 - (2\ \cdot 13) = 2 
\end{equation}
We can now observe that our end result $2$ did indeed match what we expected from the reduction.

For the hardware optimized scheme, we let $k=2$, grouping the four upper bits and their corresponding value from Table~\ref{tab:single_input_table_example} into two tables of two inputs each. The Table entries would then be:
\begin{table}[H]
    \centering
    \begin{tabular}{c|c|c}
        $[\Tilde{c}_i, \Tilde{c}_{i+1}]$ &  $f_1[\Tilde{c}_{4}, \Tilde{c}_{5}]$ &  $f_2[\Tilde{c}_{6}, \Tilde{c}_{7}]$\\ \hline
        $[0,0]$ & $0$ & $0$\\
        $[1,0]$ & $3$ & $12$\\
        $[0,1]$ & $6$ & $11$\\
        $[1,1]$ & $9$ & $(12 + 11)\mod 13 = 10$\\
    \end{tabular}
    \caption{Example of the table entries for the optimized scheme}
    \label{tab:table_optimized_example}
\end{table}
We can now again determine the $i_{\max}$ parameter, which stands for the maximum multiple of $q$ that can occur in the final subtraction step. In the optimized scheme, this is determined by the sum of the values of the lower bits in $\Tilde{c}$ that are fed into the sum directly, plus the largest entry in every table:
\begin{equation}
    \max{(\hat{c})} = \sum_{i=0}^{3} 2^i + \max{(f[\Tilde{c}_{4}, \Tilde{c}_{5}])} +\max{(f[\Tilde{c}_{6}, \Tilde{c}_{7}])}  
\end{equation}
\begin{equation*}
   = 1 + 2 + 4+ 8+ 9 +12= 36
\end{equation*}
\begin{equation}
    i_{\max} = \frac{\max{\hat{c}}}{q}=\frac{36}{13} \approx 2.8
    \label{eq:max_grouped_reduction}
\end{equation}
Applying $\Tilde{c} = 210$ to the input of this circuit will result in:
\begin{equation}
    f_1[\Tilde{c}_{4}, \Tilde{c}_{5}] = f_1[1, 0] \rightarrow 3
\end{equation}
\begin{equation*}
    f_2[\Tilde{c}_{6}, \Tilde{c}_{7}] = f_2[1,1]\rightarrow 10
\end{equation*}
\begin{equation*}
    [\Tilde{c}_{0},\Tilde{c}_{1},\Tilde{c}_{2},\Tilde{c}_{3}] = [0,1,0,0] \rightarrow 2
\end{equation*}
\begin{equation}
    \hat{c} = 2 + 3 +10 = 15
\end{equation}
\begin{equation}
     q \leq \hat{c} < 2q \rightarrow i = 1
\end{equation}
\begin{equation}
    c = \hat{c} - (i\ \cdot q) = 15 - (1\ \cdot 13) = 2 
\end{equation}
We see that we also obtain the correct result $2$ from applying this method. We can also observe that trough grouping bits together into larger LUTs, we have reduced the possible number of cases in final reduction from 3 cases in \eqref{eq:max_non_grouped_reduction} to just 2 in \eqref{eq:max_grouped_reduction}.

\section{Implementation Results}
\label{sec:impl_results}
In this section, we implement the scheme presented previously in \ref{sec:Algo_Optim_HW} on a Xilinx Zynq-7000 FPGA and present synthesis results for a potential ASIC implementation. As a modulus, we choose the moduli of the second round post quantum cryptography key encapsulation algorithm Crystals-Kyber \cite{original_kyber_paper} according to the NIST round 3 specification \cite{computer_security_division_round_2017} and the modulus of post quantum cryptography signature algorithm Crystals-Dilithium \cite{dilithium}. According to their specification, the moduli for those schemes are: $q_{\mathrm{Kyber}}=3329$ and $q_{\mathrm{Dilithium}}=8380417$. Since Kyber uses a relatively small number as a modulus and Dilithium a rather large number, testing the two can provide a good estimate on how the scheme will behave for different moduli bit-widths. The elements in $\mathbb{Z}_q$ will take $n_{\mathrm{Kyber}}=12$ and $n_{\mathrm{Dilithium}}=23$ bits. We here consider a case of reduction after the multiplication of elements in $\mathbb{Z}_q$, therefore we reduce bit vectors of size $2n$. We group the $N = n + 1$ MSBs into two LUTs for the case of Kyber (7 and 6 input bit configuration) and four 6-bit LUTs in the case of Dilithium. In FPGA implementations, we choose these groupings to make good use of the given LUT structure within a single combinatoric logic block (CLB) of the Zynq-7000 FPGA. This is at smallest a 5-bit LUT and at most a 8-bit LUT, which can be formed by grouping together 6 input LUTs. Since arithmetic operations will also be performed with LUTs in FPGAs, it is more beneficial to choose larger (within the range of the CLBs LUTs) LUT sizes rather than performing additional arithmetic operations. In an ASIC one may prefer to choose slightly smaller LUTs, but more arithmetic since table sizes can be chosen arbitrarily in ASICs and arithmetic is not built from LUTs. However, we will stick to the 6 and 7 input LUTs for the rest of this article for both ASIC and FPGA results.

As a reference, we also implement Barrett reduction from \ref{sec:barrett}, to which we will compare our results. In total, we compare 3 implementations. 'Barrett general' corresponds to the method shown in \ref{sec:barret_general}. 'Barrett optimized' corresponds to \ref{sec:barrett_optimized}. Finally, LUT based corresponds to the optimized scheme shown in \ref{sec:Algo_Optim_HW} and Fig.~\ref{fig:optimized_method}.
\subsection{FPGA Results}
Table~\ref{tab:fpga_impl_res_table} shows that the LUT-based method uses a slightly larger amount of LUTs than the general Barrett reduction, but due to the scheme not requiring any multiplications does not instantiate any DSPs. One can also see that in the optimized Barrett reduction the LUT count goes up as the DSP count goes down, since we changed multiplications to shift and adds, which is performed in the LUTs. To provide a better chip area comparison on FPGAs we also performed an implementation run with the optimized Barrett method where we do not allow the synthesis tool to use any DSPs for the multiplications. We see that here the LUT count even in the optimized scheme would be more than double compared to that of the LUT based method. This should not come as a surprise since, also multipliers/DSPs are much fewer in numbers on FPGAs  compared to LUTs, as they consume significantly more chip area per unit, which was one of the primary motivators for the LUT based scheme in the first place. For the larger bit width in Dilithium, the difference between schemes (optimized and LUT based) appear to be less significant. However, this would not generalize to arbitrary moduli of this bit width, as the modulus of Dilithium benefits heavily from having both a multiplication constant $m=8396807=2^{23} + 2^{13}+ 2^3 - 1$ and a modulus $q = 8380417 = 2^{23} - 2^{13} + 1$ that can be expressed by the sum of only very few shifted input terms to perform multiplication with them. For arbitrary moduli of this bit-width, it is therefore better to compare it to the generalized Barrett.
\begin{table}[H]
    \centering
    \begin{tabular}{c|c|c|c|c}
                                    & \multicolumn{2}{c|}{Kyber} & \multicolumn{2}{c}{Dilithium}\\
                                    & LUTs  & DSPs & LUTs  & DSPs\\ \hline
         Barrett general            & 54    & 2 & 130 & 5\\
         Barrett optimized        & 74    & 1 & 255 & 0\\
         Barrett optimized (no DSP) & 212   & 0 & - & -\\
         LUT based method           & 82    & 0 & 200 & 0
    \end{tabular}
    \caption{FPGA implementation results in terms of resource usage}
    \label{tab:fpga_impl_res_table}
\end{table}

\subsection{ASIC Synthesis Results}
We perform ASIC synthesis for four reduction variants using Cadence Genus. The purpose of this is to obtain an area comparison result, which allows for comparison in terms of chip area for implementations that both use DSPs and LUTs in an FPGA, such as the Barrett reduction. We have not further optimized the LUT sizes for ASIC implementation for this test and stayed with the FPGA optimized 6 and 7 input LUT design, which may not be the best variant for ASICs. Since the absolute chip area will depend upon the used ASIC technology standard, we only compare the area consumption relative to each other.

\begin{table}[H]
    \centering
    \begin{tabular}{c|c|c}
                                    & \multicolumn{2}{c}{Area normalized to Barrett optimized}\\ 
                                    & Kyber & Dilithium\\ \hline
         
         Barrett optimized (synthesis tool) & 1 & 1.2 (1) \\
         Barrett optimized (manual) & 0.97 &  1.2 (1)\\
         Barrett general (full multiplier)  & 2.7 & 7.4  (6.0)\\
         LUT based method           & 0.59 &  0.78 (0.63)
    \end{tabular}
    \caption{Area results of ASIC synthesis normalized to the area of the automatically optimized version of the Barrett implementation for Kyber. Terms in () are normalized within their column}
    \label{tab:asic_impl_res_table}
\end{table}

Table~\ref{tab:asic_impl_res_table} shows the result of ASIC synthesis in terms of the relative required area on a chip. The ASIC synthesis tool can perform similar optimizations to what we did manually in the optimized scheme \ref{sec:barrett_optimized}, since synthesis is performed for a specific constant modulus. The synthesis tool does not need to generate full general multipliers. We therefore also explicitly performed synthesis for the general Barrett reduction with full multipliers, this provides an estimate on how large the Barrett reduction module can become for arbitrary moduli of this bit width. It can be seen that the proposed LUT based method consumes only about 10\% to 20\%  of the chip area compared to the general Barrett reduction with full multipliers and about 60\% of the optimized Barrett reductions in the tested case.

\section{Conclusion}
In this article, we have shown a potential alternative to the commonly used Barrett reduction for computing modular reduction in hardware. The method shows promising results in terms of area consumption, which was reduced compared to optimized Barrett reduction by 39\%  for the tested modulus and even 90\% for not specifically selected arbitrary moduli. Since especially the area consumption of the optimized Barrett reduction varies greatly for different moduli, the ratio by which the scheme proposed here is better (or worse) than the Barrett method, may vary significantly depending on the modulus. For the proposed scheme itself, the area consumption will stay relatively consistent for different moduli of the same bit length. This can already be an advantage, since it removes the concern of good reducibility in the process of choosing a modulus for cryptographic schemes, which often also need to be chosen for other properties, such as being a prime number. Since all steps of the proposed method scale linearly to the bit width of the number being reduced, it may also be a viable scheme for reducing large numbers. Another major advantage of the scheme is that all operations can be performed as (close to) $n$-bit wide logic, as opposed to the operation width being determined by the input bit width. This shortens carry signal paths, allows for faster clocking of the design and fast reduction even for high bit widths. Most operations (LUTs, summation of LUT outputs) can be parallelized and are well suited for pipeling which is another useful property for hardware implementations. As a next step, one may want to further investigate and quantify the effects of different parameters such as the table size $k$, modulus $q$ or size of the number being reduced on the relative performance compared to the Barrett method as well as the total area consumption.
\AtNextBibliography{\small}
\printbibliography

@article{dilithium,
  title={Crystals-dilithium: A lattice-based digital signature scheme},
  author={Ducas, L{\'e}o and Kiltz, Eike and Lepoint, Tancrede and Lyubashevsky, Vadim and Schwabe, Peter and Seiler, Gregor and Stehl{\'e}, Damien},
  journal={IACR Transactions on Cryptographic Hardware and Embedded Systems},
  pages={238--268},
  year={2018}
}

@inproceedings{barrett_implementing_1987,
	location = {Berlin, Heidelberg},
	title = {Implementing the Rivest Shamir and Adleman Public Key Encryption Algorithm on a Standard Digital Signal Processor},
	isbn = {978-3-540-47721-1},
	doi = {10.1007/3-540-47721-7_24},
	series = {Lecture Notes in Computer Science},
	pages = {311--323},
	booktitle = {Advances in Cryptology — {CRYPTO}’ 86},
	publisher = {Springer},
	author = {Barrett, Paul},
	editor = {Odlyzko, Andrew M.},
	date = {1987},
	langid = {english}
}

@article{montgomery_modular_1985,
	title = {Modular multiplication without trial division},
	volume = {44},
	issn = {0025-5718, 1088-6842},
	url = {https://www.ams.org/mcom/1985-44-170/S0025-5718-1985-0777282-X/},
	doi = {10.1090/S0025-5718-1985-0777282-X},
	pages = {519--521},
	number = {170},
	journaltitle = {Mathematics of Computation},
	shortjournal = {Math. Comp.},
	author = {Montgomery, Peter L.},
	urldate = {2023-03-15},
	date = {1985},
	langid = {english}
}

@misc{cryptoeprint:2014/040,
      author = {Zhengjun Cao and Ruizhong Wei and Xiaodong Lin},
      title = {A Fast Modular Reduction Method},
      howpublished = {Cryptology ePrint Archive, Paper 2014/040},
      year = {2014},
      note = {\url{https://eprint.iacr.org/2014/040}},
      url = {https://eprint.iacr.org/2014/040}
}

@article{lim_fast_nodate,
	title = {Fast Modular Reduction With Precomputation},
	author = {Lim, Chae Hoon and Hwang, Hyo Sun and Jae-Dong, Yang and Cho-Gu, Seo},
	langid = {english},
        url={https://citeseerx.ist.psu.edu/document?repid=rep1&type=pdf&doi=a683d452712796ca9ab4c99dffdfd48e1e5d195d}
}

@misc{cryptoeprint:2014/755,
      author = {Mark A.  Will and Ryan K.  L.  Ko},
      title = {Computing Mod Without Mod},
      howpublished = {Cryptology ePrint Archive, Paper 2014/755},
      year = {2014},
      note = {\url{https://eprint.iacr.org/2014/755}},
      url = {https://eprint.iacr.org/2014/755}
}

@online{computer_security_division_round_2017,
	title = {Round 3 Submissions - Post-Quantum Cryptography {\textbar} {CSRC} {\textbar} {CSRC}},
	url = {https://csrc.nist.gov/Projects/post-quantum-cryptography/post-quantum-cryptography-standardization/round-3-submissions},
	author = {Computer Security Division, Information Technology Laboratory},
	urldate = {2023-03-29},
	date = {2023-03-27}
}

@article{banerjee2019sapphire,
  title={Sapphire: A configurable crypto-processor for post-quantum lattice-based protocols},
  author={Banerjee, Utsav and Ukyab, Tenzin S and Chandrakasan, Anantha P},
  journal={arXiv preprint arXiv:1910.07557},
  year={2019}
}

@INPROCEEDINGS{original_kyber_paper,
  author={Bos, Joppe and Ducas, Leo and Kiltz, Eike and Lepoint, T and Lyubashevsky, Vadim and Schanck, John M. and Schwabe, Peter and Seiler, Gregor and Stehle, Damien},
  booktitle={2018 IEEE European Symposium on Security and Privacy (EuroS\&P)}, 
  title={CRYSTALS - Kyber: A CCA-Secure Module-Lattice-Based KEM}, 
  year={2018},
  volume={},
  number={},
  pages={353-367},
  doi={10.1109/EuroSP.2018.00032}}

@INPROCEEDINGS{pqc_overview,
  author={Kumar, Manoj and Pattnaik, Pratap},
  booktitle={2020 IEEE High Performance Extreme Computing Conference (HPEC)}, 
  title={Post Quantum Cryptography(PQC) - An overview: (Invited Paper)}, 
  year={2020},
  volume={},
  number={},
  pages={1-9},
  doi={10.1109/HPEC43674.2020.9286147}}

@INPROCEEDINGS{pqc_challenges,
  author={Bavdekar, Ritik and Jayant Chopde, Eashan and Agrawal, Ankit and Bhatia, Ashutosh and Tiwari, Kamlesh},
  booktitle={2023 International Conference on Information Networking (ICOIN)}, 
  title={Post Quantum Cryptography: A Review of Techniques, Challenges and Standardizations}, 
  year={2023},
  volume={},
  number={},
  pages={146-151},
  doi={10.1109/ICOIN56518.2023.10048976}}

@incollection{r_omondi_modular_2020,
	location = {Cham},
	title = {Modular Reduction},
	isbn = {978-3-030-34142-8},
	url = {https://doi.org/10.1007/978-3-030-34142-8_4},
	pages = {105--141},
	booktitle = {Cryptography Arithmetic: Algorithms and Hardware Architectures},
	publisher = {Springer International Publishing},
	author = {R. Omondi, Amos},
	date = {2020},
	doi = {10.1007/978-3-030-34142-8_4},
}

\end{document}